\documentclass[11pt,a4paper]{article}
\pdfoutput=1
\usepackage{jcappub}
\usepackage{slashed}

\def\be{\begin{equation}}
\def\ee{\end{equation}}
\def\bea{\begin{eqnarray}}
\def\eea{\end{eqnarray}}

\begin{document}

\title{Dual Path Integral\\\LARGE a non-perturbative approach to strong coupling}

\author{Vitaly Vanchurin}

\emailAdd{vvanchur@d.umn.edu}

\date{\today}

\affiliation{Department of Physics, University of Minnesota, Duluth, Minnesota, 55812 \\
Duluth Institute for Advanced Study, Duluth, Minnesota, 55804}

\abstract{We develop a non-perturbative method for calculating partition functions of strongly coupled quantum mechanical systems with interactions between subsystems described by a path integral of a dual system. The dual path integral is derived starting from non-interacting subsystems at zeroth order and then by introducing couplings of increasing complexity at each order of an iterative procedure. These orders of interactions play the role of a dual time and the full quantum partition function is expressed as a transition amplitude in the dual system. More precisely, it is expressed as a path integral from a deformation-operators dependent initial state at zero time/order to the inverse-temperature dependent final state at later time/order. We provide examples of strongly coupled systems with up to first-order interactions (e.g. Ising model) and arbitrary high-order interactions (e.g. $1+1$D QFT). We also discuss a possible emergence of space-time, quantum field theories and general relativity in context of the dual path integral. }

\maketitle

\section{Introduction}

There are two main puzzles in theoretical physics which so far have no satisfactory resolutions. The first one is how to calculate observables in strongly coupled theories and the second one is how to derive general relativity from quantum mechanics, i.e. the problem of quantum gravity. While the former puzzle is purely mathematical, as it is usually well defined although the analytical calculations can be difficult to carry out, the latter puzzle is more physical than mathematical, as it is not even clear what is the right problem to solve. Of course, it would be nice if the two problems were actually related and in fact there are evidences suggesting that it might be the case. For example, in context of the AdS/CFT correspondence \cite{Maldacena, Witten}, general relativity in the bulk emerges from a quantum theory on the boundary and at the same time observables of a strongly coupled conformal field theory can be obtained by performing perturbative calculations on the anti-de Sitter background. Unfortunately, the space-time we live in is not anti-de Sitter and so the AdS/CFT conjecture is at most an educated guess of what one might expect to see in a true theory of quantum gravity.

In this paper our main motivation is to study the emergence of space-time and general relativity from strongly coupled theories beyond the AdS/CFT duality, but the initial task will be a lot more modest. What we want first is to understand how a fully interacting quantum mechanical partition function can be obtained from the partition functions of non-interacting subsystems. In Ref. \cite{Vanchurin1} we constructed one such transformation described by a second order differential equation for the ``child'' partition function with initial conditions specified by a ``parent'' partition function. In certain cases the parent partition function can describe non-interacting subsystems (see Sec. \ref{sec:spacetime}), but the framework was not rich enough to handle more complex systems including many strongly coupled systems. In this paper we will extend the analysis to a lot more general transformations (and thus more general interactions) described by a path integral with initial state specified by non-interacting partition functions (see Sec. \ref{sec:path_integral}). By construction, the analysis is non-perturbative with couplings of different orders of complexity introduced at different orders of an iterative procedure.

At a very minimum, the dual description can be viewed as a non-perturbative method for calculating partition functions of strongly coupled quantum mechanical systems such as spin chain models (see Sec. \ref{sec:ising}), but more ambitiously it may also tell us something new about the quantum origin of space-time and gravity. In Ref. \cite{Vanchurin2} we argued that the anti-commutativity of quantum operators may be responsible for the emergence of space-time and in this paper we see that essentially the same phenomena occurs for more general quantum systems (see Sec. \ref{sec:spacetime}). Of course, it is premature to claim that we have derived the fully non-perturbative equations of general relativity (see, however, some promising recent ideas in context of emergent gravity \cite{Jacobson, Verlinde, Vanchurin3, ML, NN}), but some progress in this directions can be made. In particular, we will argue that the fully interacting quantum partition function can often be represented by a dual path integral of a local quantum filed theory  (see Sec. \ref{sec:qft}) where the perturbative methods are of value. With this respect the proposed framework can be considered as a generalization of the AdS/CFT duality mapping to more general quantum systems. Even more generally, the dual path integral may differ from the Feynman path integral and then some significant deviations from the quantum field theory framework are expected.

The paper is organized as follows.  In the next section we construct a quantum system with a particular type of interactions which can be modeled with an additional (or first order) system. In Sec. \ref{sec:spacetime} we consider a simple system with first-order interactions which can be described through a transformation of an extended partition function in an emergent space-time. In Sec. \ref{sec:continuation} we obtain an analytically continued expression of the the partition function which is used in  Sec. \ref{sec:path_integral} to derive the dual path integral representation of a fully interacting partition function. In Sec. \ref{sec:qft} we argue that the local quantum field theory may emerge in certain limits of the dual path integral, but some significant deviations from the Feynman path integral are also expected. In Secs. \ref{sec:spin} and \ref{sec:ising} we demonstrate how the dual description of spin chain models can be used to study strongly coupled systems with respectively local and non-local interactions. In Sec. \ref{sec:discussion} we summarize and discuss the main results of the paper, i.e. the dual path integral and emergent phenomena. 

\section{First-order interactions} \label{sec:interactions}

Consider a quantum system described by a Hamiltonian operator
\be
\hat{H}_0 \equiv \sum_{k=1}^K  J^{0}_k \hat{H}^{k}_0 \label{eq:Hamiltonian}
\ee
where $J^{0}_k$'s are the real coefficients which parametrize deformations of the corresponding Hermitian operators $\hat{H}^{k}_{0}$'s. At finite temperature the system is described by the quantum partition function
\be
{\cal  Z}[ - \beta, J ] = tr\left[\exp\left(- \beta \sum_{k=1}^K J^{0}_k  \hat{H}^{k}_0   \right ) \right ]\label{eq:partition_function}
\ee
where $\beta$ is the inverse temperature. To simplify calculations in Sec. \ref{sec:continuation}  we shall analytically continue the parameter $\beta$ to complex plane, but it will be assumed throughout the paper that we are dealing with a quantum system in thermal equilibrium. The time will eventually emerge in the dual descriptions first as a relativistic coordinate in Sec. \ref{sec:spacetime} and later as a parameter in a path integral of Sec. \ref{sec:path_integral} despite of the fact that the original quantum system is in a time-invariant thermal state.

In general the partition function \eqref{eq:partition_function} is difficult calculate, but the analysis if greatly simplified if the operators $\hat{H}^{k}_0$ describe Hamiltonian operators of non-interacting subsystems.  This happens when the entire Hilbert space can be decomposes into a tensor product of Hilbert spaces 
\be
{\cal H}_0 = {\cal H}^1_0 \otimes {\cal H}^2_0 ... \otimes {\cal H}^K_0\label{eq:Hilbert1}
\ee
 with each of the operators $\hat{H}^{k}_{0}$ acting non-trivially on the corresponding factor ${\cal H}^k_0$ only.  Then the full partition function can be expanded into a product, 
\bea
{\cal  Z}_0[-\beta, J ] &=& tr_0\left[\exp\left( - \beta \sum_{k=1}^K J^{0}_k \hat{H}^{k}_0    \right ) \right ]\notag\\
 &=& \prod_{k=1}^K  tr_0^k\left[\exp\left (- \beta J^{0}_k  \hat{H}^{k}_0 \right ) \right ] \notag\\
 &=& \prod_{k=1}^K{\cal  Z}_0^{k}[-\beta J_k] \label{eq:non_interacting_function},
\eea
where $tr_0^k[ \;]$ is the trace over $k$'th subsystem and ${\cal  Z}_0^{k}[-\beta J_k]$ is the partition function of the $k$'th subsystem.

Perhaps, more realistically, the Hamiltonian operator should contain interactions between subsystems. For example, a tensor product operator $\hat{H}^{j}_0 \otimes \hat{H}^{k}_0$ acts non-trivially on two Hilbert space factors ${\cal H}^j_0$ and ${\cal H}^k_0$, but an inclusion of such operators would spoils the factorization of the partition function and could lead to the problem of strong coupling. In this paper we will take a somewhat different approach and to describe interactions between zeroth-order subsystems ${\cal H}^k_0$'s we will employ  a collection of additional (or the first-order) operators $\hat{H}_1^1, ..., \hat{H}_1^K$ which act non-trivially on an additional (or the first-order) system ${\cal H}_1$ only. Then the full Hilbert space is a tensor product of the zeroth-order and first-order Hilbert spaces, i.e.
\be
{\cal H}_1 \otimes {\cal H}_0 = {\cal H}_1 \otimes  {\cal H}^1_0 \otimes {\cal H}^2_0 ... \otimes {\cal H}^K_0.\label{eq:Hilbert2}
\ee
and an interacting Hamiltonian can be defined as
\be
\hat{H} = \sum_{k=1}^K   \hat{H}^k_1 \otimes \hat{H}^{k}_0  J^{0}_k .\label{eq:Hamiltonian2}
\ee
This is certainly not the most general operator which can be defined in the Hilbert space \eqref{eq:Hilbert2}, but it is general enough to enable us to study certain strongly coupled systems whose partition function can be expressed through a dual path integral.

Different operators $\hat{H}^k_1 \otimes \hat{H}^{k}_0$ in \eqref{eq:Hamiltonian2} act non-trivially on the same factor of the Hilbert space, namely ${\cal H}_1$, and as a result the interacting partition function  
\be
{\cal  Z} [- \beta, J ] = tr\left[\exp\left(- \beta \sum_{k=1}^K   \hat{H}^k_1 \otimes \hat{H}^{k}_0   J^{0}_k \right ) \right ] \label{eq:interacting_function}
\ee
does not factorize as was the case for non-interacting partition function \eqref{eq:non_interacting_function}. To derive a useful expression of the interacting partition function it will be convenient to first define the first-order partition function 
\be
{\cal Z}_1[- \beta , p_1, ..., p_K] = tr_1\left [\exp\left (- \beta\sum_{k=1}^K \hat{H}_1^k p_k \right )\right ] \label{eq:combiner}
\ee
where $tr_1[ \;]$ is a trace over only the first-order system ${\cal H}_1$. The key observation is that for the system described by \eqref{eq:Hamiltonian2} the interacting partition function \eqref{eq:interacting_function} can be expressed as 
\be
{\cal  Z}[- \beta, J ] =  \left. tr_1\left [ \exp \left(-\beta \sum_{k=1}^K\hat{H}^k_1  \frac{\partial}{\partial x_0^k} \right)\right ]  \prod_{k=1}^K{\cal  Z}^{k}_0[x_0^k, J_{k}] \right |_{x_0^1=...x_0^K=0}
\ee
or in terms of the first-order partition function \eqref{eq:combiner} as
\be
{\cal  Z}[- \beta, J ] =  \left. {\cal Z}_1\left [-\beta, \frac{\partial}{\partial x_0^1}, ..., \frac{\partial}{\partial x_0^K}\right ]   \prod_{k=1}^K{\cal  Z}^{k}_0[x_0^k, J_{k}] \right |_{x_0^1=...x_0^K=0}.\label{eq:interacting_function1}  
\ee
Note that such a representation of the partition function with operator $ {\cal Z}_1\left [-\beta, \frac{\partial}{\partial x_0^1}, ..., \frac{\partial}{\partial x_0^K}\right ]$ acting on a non-interacting partition function $\prod_{k=1}^K{\cal  Z}^{k}_0[x_0^k, J_{k}]$   was only possible because the zeroth-order subsystems ${\cal H}^k_0$'s were not coupled to each other directly, but  through interactions in the first-order system as is evident from the Hamiltonian expression \eqref{eq:Hamiltonian2}.

\section{Extended partition function}\label{sec:spacetime}

In order to better understand the transformation \eqref{eq:interacting_function1} from a non-interacting partition function $\prod_{k=1}^K{\cal  Z}^{k}_0[x_0^k, J_{k}]$   to an interacting partition function partition function ${\cal  Z}[- \beta, J_1, ..., J_K]$, consider a first-order system described by anti-commuting first-order operators
\be
\{ \hat{H}^j_1, \hat{H}^k_1 \} = 2 \delta_{jk} \hat{I}_1, \label{eq:anticommutator}
\ee
where $\hat{I}_1$ is the identity operator in  ${\cal H}_1$. The corresponding first-order partition function can be calculated by separating odd and even terms in the Taylor series expansion, i.e.
\bea
{\cal Z}_1[- \beta , p_1, ..., p_K]  &=& tr_1 \left ( \exp \left (- \beta \sum_{k=1}^K \hat{H}_1^k p_k\right ) \right ) \label{eq:anticomm_partition0}\\
&=& \sum_{n=0}^\infty tr_1 \left [ \frac{(- \beta)^n}{n!}  \left ( \sum_{k=1}^K \hat{H}_1^k p_k\right )^n\right ) \notag\\
&=& \sum_{m=0}^\infty tr_1 \left ( \frac{(- \beta)^{2m}}{(2m)!}  \left ( \sum_{k=1}^K \hat{H}_1^k p_k\right )^{2m}\right ) + \sum_{m=0}^\infty tr_1 \left ( \frac{(- \beta)^{2m+1}}{(2m+1)!}  \left ( \sum_{k=1}^K \hat{H}_1^k p_k\right )^{2m+1} \right )  \notag
\eea
From the anti-commutation relation \eqref{eq:anticommutator} we get 
\be
\left (  \sum_{k=1}^K \hat{H}_1^k p_k \right )^2 = \sum_{k=1}^K\hat{I}_1 p_k^2 
\ee
which can be substituted into \eqref{eq:anticomm_partition0},
\be
{\cal Z}_1[- \beta , p_1, ..., p_K]   = \sum_{m=0}^\infty tr_1 \left ( \frac{(- \beta)^{2m}}{(2m)!}  \left ( \sum_{k=1}^K\hat{I}_1 p_k^2 \right )^{m}\right ) + \sum_{m=0}^\infty tr_1 \left ( \frac{(- \beta)^{2m+1}}{(2m+1)!}  \left ( \sum_{k=1}^K\hat{I}_1 p_k^2\right )^{m} \left ( \sum_{k=1}^K \hat{H}_1^k p_k\right ) \right ).\label{eq:anticomm_partition}
\ee
If we also assume that the anti-commuting operators are traceless 
\be
tr_1\left ( \hat{H}^j_1\right ) = 0\label{eq:traceless}
\ee
then the second term  in \eqref{eq:anticomm_partition}  vanishes and the partition function is greatly simplified 
\bea
{\cal Z}_1[- \beta , p_1, ..., p_K]   &=& tr_1 \left (\hat{I}_1\right )  \sum_{m=0}^\infty  \frac{(- \beta)^{2m}}{(2m)!}  \left ( \sum_{k=1}^K  p_k^2  \right )^{m} \notag\\
& =& {\cal N}_1 \cosh\left (\beta \sqrt{\sum_{k=1}^K p_k^2 } \right )\label{eq:anticomm_partition2}
\eea
where ${\cal N}_1$ is the trace of the identity operator $\hat{I}_1$ in the first-order system ${\cal H}_1$ (see Ref. \cite{Vanchurin2} for details).

By combining \eqref{eq:anticomm_partition2} and \eqref{eq:interacting_function1} we obtain an equation for a fully interacting partition function 
\be
{\cal  Z}[- \beta, J ] =  \left. {\cal N}_1\cosh \left (\beta \sqrt{\sum_{k=1}^K \left( \frac{\partial}{\partial x_0^k} \right)^2} \right)   \prod_{k=1}^K{\cal  Z}^{k}_0[x_0^k, J_{k}] \right |_{x_0^1=...x_0^K=0} \label{eq:interacting_function2} 
\ee
which can be calculated by following the analysis of Ref. \cite{Vanchurin2}. Let us define an extended partition function
\bea
z[- \beta, x^1, ..., x^K, J_1, ..., J_K] &\equiv&  tr\left [ \exp \left(-\beta \sum_{k=1}^K \hat{H}^k_1  \frac{\partial}{\partial x_k} \right)\right ]  \prod_{k=1}^K{\cal  Z}^{k}_0[x^k, J_{k}] \notag\\
& = & {\cal N}_1 \cosh \left (\beta \sqrt{\sum_{k=1}^K \left (\frac{\partial}{\partial x^k}\right )^2  } \right ) \prod_{k=1}^K{\cal  Z}^{k}_0[x^k, J_{k}]\label{eq:extended_function}
\eea
that contains information about both interacting and non-interacting partition functions. For example, at $\beta=0$ the function reduces to a product of non-interacting partition functions of the zeroth-order subsystems, i.e.
\be
z[0, x^1, ..., x^K, J_1, ..., J_K] = {\cal N}_1  \prod_{k=1}^K{\cal  Z}^{k}_0[x^k, J_{k}] \label{eq:initial_condition1}
\ee
and at $x^1 = ... = x^K=0$ the function reduces to the fully interacting partition function, i.e.
\be
z[-\beta, 0, ..., 0, J_1, ..., J_K] = {\cal  Z}[- \beta, J_1, ..., J_K]. \label{eq:final_condition}
\ee
Then one might wonder if it may be possible to derive a differential equation whose solution would be the entire extended partition function (including \eqref{eq:final_condition}) starting from initial conditions at $\beta=0$, i.e. from \eqref{eq:initial_condition1}. 

The answer to the question is affirmative and the corresponding equation can be found by differentiating (twice) the extended partition function \eqref{eq:extended_function} with respect to the inverse temperature, i.e.
\bea
\frac{\partial^2}{\partial \beta^2} z[- \beta, x, J] &=& \left ( \sum_{k=1}^K \left (\frac{\partial}{\partial x^k}\right )^2 \right ){\cal N}_1 \cosh \left (\beta \sqrt{\sum_{k=1}^K \left (\frac{\partial}{\partial x^k}\right )^2  } \right ) \prod_{k=1}^K{\cal  Z}^{k}_0[x^k, J_{k}]\notag\\
&=& \left ( \sum_{k=1}^K \left (\frac{\partial}{\partial x^k}\right )^2 \right ) z[- \beta, x, J].
\eea
Then it is obvious  that the extended partition function must be a solution of a relativistic wave equation
\be
\Box z[\beta, x, J]  = \left ( \frac{\partial^2}{\partial \beta^2} -  \sum_{k=1}^K \left (\frac{\partial}{\partial x^k}\right )^2\right ) z[\beta, x, J]  =0, \label{eq:wave}
\ee
but since \eqref{eq:wave} is a second order differential equation in $\beta$, the correct initial conditions must be specified at $\beta=0$ for both``position'' $z[0, x, J]$ and ``velocity'' $\frac{\partial}{\partial \beta}z[0, x, J]$. These conditions can be deduced directly from the definition \eqref{eq:extended_function}, i.e.
\bea
z[0, x, J] &=&{\cal N}_1 \prod_{k=1}^K{\cal  Z}^{k}_0[x^k, J_{k}]\\
\frac{\partial}{\partial \beta}z[0, x, J] &=& 0. 
\eea
Evidently, equation \eqref{eq:wave} describes a transformation from the initial state $z[0, x, J]$ specified by a non-interacting partition function to final state $z[\beta, x, J]$ which describes the fully interacting partition function $z[\beta, 0, J]=  {\cal  Z}[- \beta, J]$ at inverse temperature $\beta$.

In the remainder of the paper we shall generalize this transformation to a lot more general couplings between subsystems, but before we proceed let us emphasize that the transformation is described by an equation which is  relativistic \eqref{eq:wave}. Recall that the original quantum system was in a time-invariant thermal state and it is only in the dual description the relativistic dynamics of the extended partition function emerged from a particular coupling between subsystems. 

\section{Analytic continuation} \label{sec:continuation}

So far, we have analyzed some very specific first-order interactions between zeroth-order subsystems described by a quantum Hamiltonian \eqref{eq:Hamiltonian2}. This allowed us to  derive  the corresponding interacting partition function \eqref{eq:interacting_function1} and to study a possible emergence of  space-time \eqref{eq:wave}. The next step is to generalize the construction to allow more general interactions between subsystems and then (if we are lucky) to study a possible emergence of  general relativity from strong coupling. In the following section we will describe one such generalization by including the higher-order interactions, but it turns out that such interactions are a lot more transparent when expressed in terms of the analytically continued functions. 

Consider the quantum partition function \eqref{eq:interacting_function} analytically continued to complex plane,\footnote{The parameter $ix$ is purely imaginary, but it is assumed that at the end of the calculations it is to be analytically continued back to the negative real line, i.e. $-\beta$.}
\be
{\cal  Z}[ix, J ] = tr\left[\exp\left(ix \sum_{k=1}^K   \hat{H}^k_1 \otimes \hat{H}^{k}_0   J^{0}_k \right ) \right ].
\ee
This function was shown to be given by the equation \eqref{eq:interacting_function1} which can also be written using the Dirac delta functions,
\bea
{\cal  Z}[ix , J] &=& \int \left ( \prod_{k=1}^K d x_0^k \; \delta(x_0^k) \right )    {\cal Z}_1\left [i x, \frac{\partial}{\partial x_0^1}, ..., \frac{\partial}{\partial x_0^K}\right ]  \prod_{k=1}^K{\cal  Z}^{k}_0[x_0^k, J_{k}]  \notag\\
 &=& \int \left ( \prod_{k=1}^K d x_0^k \; \delta(x_0^k) \right )    tr_1\left [ \exp \left(x  \sum_{k=1}^K\hat{H}_1^{k}  \frac{\partial}{\partial x_0^k} \right)\right ]   \prod_{k=1}^K{\cal  Z}^{k}_0[i x_0^k, J_{k}]  \notag\\
&=& \int \left ( \prod_{k=1}^K \frac{d x_0^k d p_k^1}{2 \pi}  e^{- i  x_0^k p_k^1}    \right )    tr_1\left [ \exp \left(x \sum_{k=1}^K\hat{H}_1^{k}  \frac{\partial}{\partial x_0^k} \right)\right ] \prod_{k=1}^K{\cal  Z}^{k}_0[i x_0^k, J_{k}].
\eea
Upon integration by parts and neglecting the (vanishing) boundary terms we get
\bea
{\cal  Z}[ix , J]  &=& \int \prod_{k=1}^K\frac{d x_0^k d p_k^1}{2 \pi}  \prod_{k=1}^K {\cal  Z}^{k}_0[i x_0^k, J_{k}]   \;  tr_1\left [ \exp \left(- x  \sum_{k=1}^K\hat{H}_1^{k}  \frac{\partial}{\partial x_0^k} \right)\right ]  e^{- i \sum_{k=1}^K x_0^k p_k^1}   \notag\\
&=& \int \prod_{k=1}^K \frac{d x_0^k d p_k^1}{2 \pi}  \prod_{k=1}^K  {\cal  Z}^{k}_0[ix_0^k, J_{k}]   \;  tr_1\left [ \exp \left( i x  \sum_{k=1}^K p_k^1  \hat{H}_1^{k} \right)\right ]   e^{- i \sum_{k=1}^K x_0^k p_k^1}  \notag\\
&=& \int  \prod_{k=1}^K\frac{d x_0^k d p_k^1}{2 \pi}    \prod_{k=1}^K  {\cal  Z}^{k}_0[ix_0^k, J_{k}] \; e^{- i\sum_{k=1}^K x_0^k p_k^1}\;  {\cal Z}_1[i x, p^1_1, ..., p^1_K]    \label{eq:combiner_transfromaton}.
\eea
Just like before \eqref{eq:interacting_function1} the new equation \eqref{eq:combiner_transfromaton} describes a transformation from a zeroth-order partition function of non-interacting subsystems $\prod_{k=1}^K  {\cal  Z}^{k}_0[ix_0^k, J_{k}]$ (in the Hilbert space ${\cal H}_0$) to a fully interacting partition function  ${\cal  Z}[ix , J]$ (in the Hilbert space ${\cal H}_1 \otimes {\cal H}_0$) with interactions described by the first-order partition function ${\cal Z}_1[i x, p^1_1, ..., p^1_K]$. For example, the analytically continued partition function \eqref{eq:interacting_function2} can now be written as
\be
{\cal  Z}[ix , J] = \int  \prod_{k=1}^K\frac{d x_0^k d p_k^1}{2 \pi}    \prod_{k=1}^K  {\cal  Z}^{k}_0[ix_0^k, J_{k}] \; e^{- i\sum_{k=1}^K x_0^k p_k^1}\;   {\cal N}_1\cos \left (x \sqrt{\sum_{k=1}^K \left( p^1_k \right)^2} \right)      \label{eq:example1}.
\ee 

The key idea is, whenever possible, to expand  ${\cal Z}_1[i x, p^1_1, ..., p^1_K]$ into a product of the first-order non-interacting partition functions  ${\cal Z}_1^j[ix,   p_{1}, ..., p_{K} ]$  similarly to how the zeroth-order partition function was expanded into a product of the zeroth-order non-interacting partition functions \eqref{eq:non_interacting_function}. Then the combined first-order partition function is
\bea
{\cal Z}_1[i x,  p_{1}, ..., p_{K}] &=& tr\left [\exp\left (ix \sum_{j=1}^K \sum_{k =1}^K \hat{H}_1^{j k}  p_{k}  \right )\right ] \notag \\
&=&   \prod_{j=1}^K  tr_1^j\left [\exp\left (ix \sum_{k=1}^K  \hat{H}_1^{j k}  p_{k} \right )\right ]  \notag\\
&\equiv&  \prod_{j=1}^K {\cal Z}_1^j [ix, p_{1}, ..., p_{K}].\label{eq:first_order0}
\eea
Such a factorization is possible when the first-order Hilbert space can be decomposed into a tensor product ${\cal H}_1 = {\cal H}^1_1 \otimes {\cal H}^2_1 ... \otimes {\cal H}^K_1$  with operators $\hat{H}^{jk}_{1}$ acting non-trivially on only a single factor ${\cal H}^j_1$ for all $k$.  To simplify calculations, but without loosing the generality, we assume that the number of factors in the first-order Hilbert space is once again $K$ and the full Hilbert space is 
\be
{\cal H} = {\cal H}_1 \otimes {\cal H}_0 = \left ( \bigotimes_{j=1}^K {\cal H}_1^j \right ) \left( \bigotimes_{k=1}^K {\cal H}_0^k\right).
\ee
The factorization of the first-order partition function \eqref{eq:first_order0} can be substituted into \eqref{eq:combiner_transfromaton} to obtain
\bea
{\cal  Z}[ix, J]   &=& tr_1\left[\exp\left(ix \sum_{k=1}^K \left ( \sum_{j=1}^K   \hat{H}_1^{j k} \right )  \otimes  \hat{H}^{k}_0   J^{0}_k \right ) \right ]  \label{eq:first_order}\\
&=& \int  \prod_{k=1}^K\frac{d x_0^k d p_k^1}{2 \pi}    \prod_{k=1}^K  {\cal  Z}^{k}_0[i x_0^k, J_{k}] \; e^{- i\sum_{k=1}^K x_0^k p_k^1}\; \prod_{j=1}^K {\cal Z}^j_1[ix, p^1_1, ..., p^1_K]. \notag
\eea
This is the most general partition function with interactions between zeroth-order subsystems ${\cal H}^k_0$'s described by the first-order subsystems ${\cal H}^k_1$'s which are not yet interacting.

\section{Dual path integral} \label{sec:path_integral}

To  derive a path integral with higher-order interactions all that we have to do is to iterate the procedure developed above. For example, the second-order interactions between the first-order subsystems ${\cal H}^k_1$'s can be introduced by employing a second-order system ${\cal H}_2$ which (for starters) can be assumed to factor into a tensor product of non-interacting second-order subsystems  ${\cal H}^k_2$'s. Then the full Hilbert space consists of a tensor product of the zeroth-, first- and second-order systems, i.e.
\be
{\cal H} = {\cal H}_2 \otimes {\cal H}_1 \otimes {\cal H}_0 = \left ( \bigotimes_{l=1}^K {\cal H}_2^l \right ) \left ( \bigotimes_{j=1}^K {\cal H}_1^j \right ) \left( \bigotimes_{k=1}^K {\cal H}_0^k\right).
\ee
By analogy with the first-order operators we define the second-order operators $\hat{H}^{lj}_2$ which act non-trivially on only a single factor ${\cal H}^l_2$ of the Hilbert space. Then the full partition function can be obtained trivially from \eqref{eq:first_order} by adding the second order of interactions, i.e.
\bea
{\cal  Z}[ix, J]   &=& tr\left[\exp\left(ix \sum_{l=1}^K \sum_{j=1}^K  \sum_{k=1}^K  \hat{H}_2^{l j}  \otimes  \hat{H}_1^{j k}  \otimes \hat{H}^{k}_0   J^{0}_k \right ) \right ] \label{eq:second_order}  \\
&=& \int  \prod_{k=1}^K\frac{d x_0^k d p_k^1d x_1^k d p_k^2}{(2 \pi)^2}    \prod_{k=1}^K  {\cal  Z}^{k}_0[i x_0^k, J_{k}] \; e^{- i\sum_{k=1}^K x_0^k p_k^1}\; \prod_{k=1}^K {\cal Z}^k_1[ix^k_1, p^1] e^{- i\sum_{k=1}^K x_1^k p_k^2} \; \prod_{k=1}^K {\cal Z}^k_2[ix, p^2].  \notag
\eea
This is now the partition function with up to the second order of interactions complexity, but we can keep going and introduce third-order interactions, fourth-order interactions etc. 

To describe a fully interacting partition function (with up to some finite order $N$ of interactions) we define a tensor product Hilbert space
\be
{\cal H} =\bigotimes_{n=0}^N  {\cal H}_n = \bigotimes_{n=0}^N   \bigotimes_{k=1}^K {\cal H}_n^k.
\ee
Consequently, the operators $\hat{H}_n^{jk}$ are assumed to act non-trivially on only respective Hilbert space factors ${\cal H}_n^j$. From this point on it will be useful to borrow some terminology from the textbook quantum mechanics, e.g. propagator, wave-functions, Hamiltonian, path integral, etc., although as we shall see the analogy is not exact. To avoid confusions and also to distinguish the approximate notions of the dual description from the standard terminology we shall be adding the word ``dual'' wherever appropriate, e.g. dual propagator, dual wave-functions, dual Hamiltonian, dual path integral, etc.

For example, it will be convenient to define a ``dual Hamiltonian density'' as
\be
 {H}_n^j(x,  p_1, ..., p_K )  \equiv x p_j + i \log \left ({\cal  Z}^{j}_n[i x, p_{1}, ..., p_{K}] \right ).\label{eq:Hamiltonian_density}
\ee
where the $n$'th order partition functions of non-interacting subsystems are given by
\be
{\cal  Z}^{j}_n[i x, p_{1}, ..., p_{K}] =  tr_n^j\left [\exp\left (i x \sum_{k=1}^K  \hat{H}_n^{j k}  p_{k} \right )\right ]
\ee
and  $tr_n^j[ \;]$ is a trace over ${\cal H}_n^j$ factor. Then the combined partition function of all $n$'th order subsystems can be written as
\bea
{\cal Z}_n[ix,  p_{1}, ..., p_{K}] &=& tr_n\left [\exp\left (ix \sum_{j=1}^K \sum_{k =1}^K \hat{H}_n^{j k}  p_{k}  \right )\right ] \notag \\
&=&   \prod_{j=1}^K  tr_n^k\left [\exp\left (i x \sum_{k=1}^K  \hat{H}_n^{j k}  p_{k} \right )\right ] \notag\\ 
&=& \prod_{j=1}^K {\cal Z}_n^j [i x, p_{1}, ..., p_{K}]\notag \\
&=& \exp\left ( \sum_{j=1}^K \left (i x  p_{j} - i {H}_n^j(x, p_{1}, ..., p_{K})  \right ) \right )
\eea
where $tr_n[ \;]$ is a trace over ${\cal H}_n = \bigotimes_{k=1}^K {\cal H}_n^k$. At this point it is not clear why \eqref{eq:Hamiltonian_density} should be called the dual Hamiltonian density, but as we shall see shortly the choice is very well motivated.  

\begin{figure}[]
\begin{center}
\includegraphics[width=0.9\textwidth]{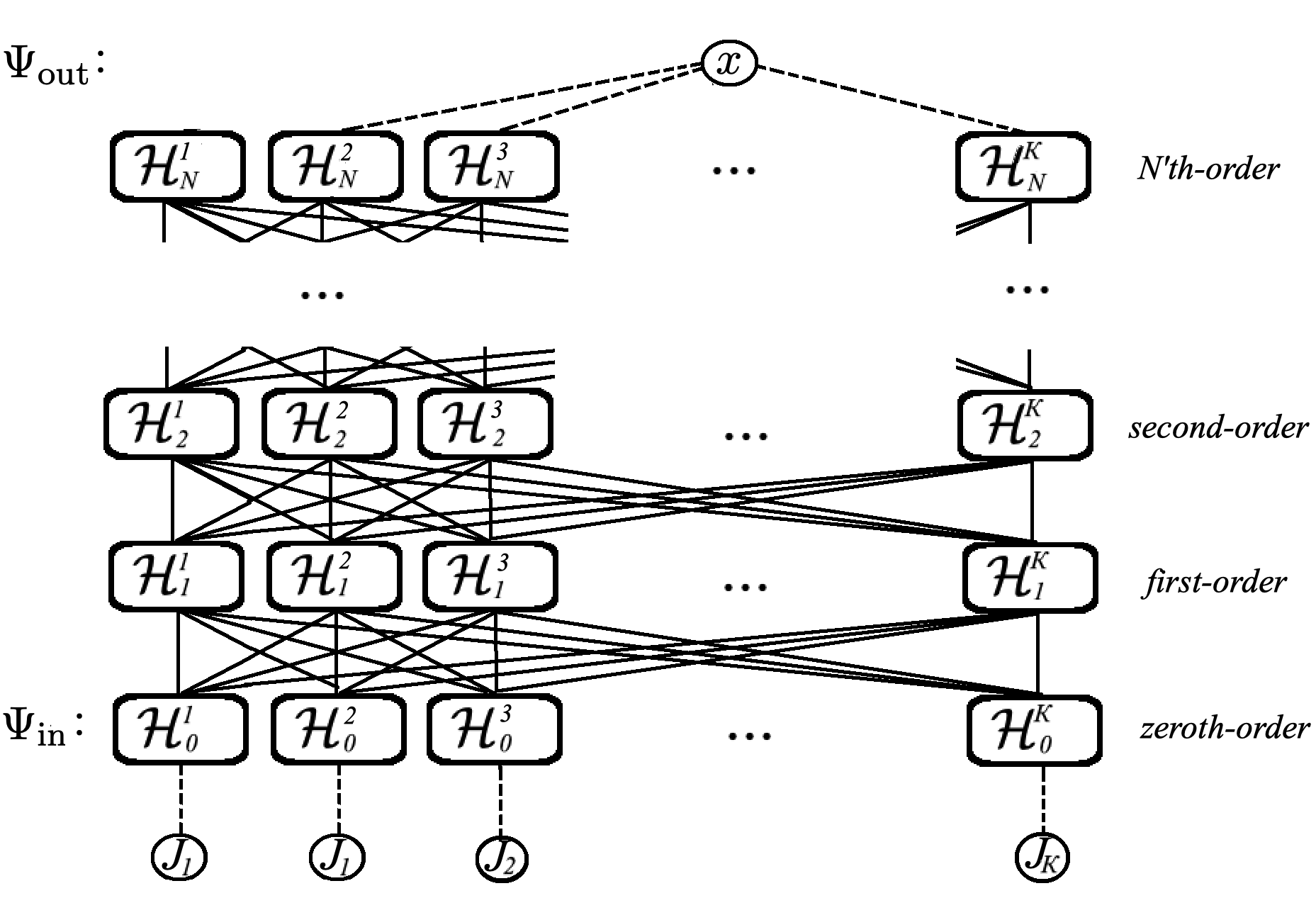}
\caption{Interactions network of a dual path integral.} \label{fig:network}
\end{center}
\end{figure}

To obtain a ``dual path integral'' expression for the fully interacting partition function we iterate the procedure which had led to \eqref{eq:second_order}. The final result is
\be
{\cal  Z}[i x, J] \equiv  \int d^K x_0 d^K x_N\;\Psi_\text{out}[x_N] {\cal K}[x_N; x_0] \Psi_\text{in}[x_0] \label{eq:path_integral}
\ee
where the ``dual propagator'' is defined as
\be
{\cal K}[x_N;x_0]   =  \int \prod_{k=1}^K  \frac{ d p_k^1 d x^k_{1}  ... d x^k_{N-1} d p_k^N }{(2 \pi)^{N}}  \; e^{- i  \sum_{n=1}^{N} \sum_{k=1}^K \left ( {H}^k_n( x^k_{n},  p^n_1, ..., p^n_K) -  \left ( x^k_{n} -x^k_{n-1} \right )p_k^{n} \right )}
\ee
 and the initial and final ``dual wave-functions'' as
\bea
\Psi_\text{in}[x_0^1, ... ,x_0^K]  &\equiv&\prod_{k=1}^K {\cal Z}_0^k(x_0^k, J_k)  \label{eq:initial_state}\\
\Psi_\text{out}[x_N^1, ..., x_N^K]  &\equiv&\prod_{k=1}^K \delta(x-x_N^k) .\label{eq:final_state}.
\eea
See Fig. \ref{fig:network} for an illustration of interactions in a generic dual path integral with boxes representing separate factors of the Hilbert space ${\cal H}_n^k$, solid lines representing tensor products of operator from the respective factors and dashed lines connected to circles representing variables of the full partition function ${\cal  Z}[i x, J]$, i.e. $x$ and $J_1, ..., J_K$.
Note that the dual path integral \eqref{eq:path_integral} can be interpreted as a transition amplitude of a dual system from the deformation-operators dependent initial state \eqref{eq:initial_state} to inverse-temperature dependent final state \eqref{eq:final_state}. This in a sharp contrast to the standard path integral representation of thermal partition functions where the inverse temperature plays the role of the size of extra dimension (see for example Refs. \cite{KirznitsLinde, Weinberg, Jackiw}).

\section{Quantum field theories}\label{sec:qft} 

The dual path integral expression \eqref{eq:path_integral} is exact, but it involves calculating a transition amplitude from initial state \eqref{eq:initial_state} to final state  \eqref{eq:final_state}  in a system described by a ``dual time''-dependent (or $n$-dependent) and ``dual space''-dependent (or $k$-dependent) Hamiltonian density ${H}^k_n (x^k,  p_1, ..., p_K)$. This can be considered as a generalization of the Feynman path integral for local quantum field theories where the Hamiltonian density is usually assumed to be the same everywhere in space and time. However, for more general geometries with dynamical gravitational degrees of freedom the Feynman path integral is not adequate and with this respect the dual path integral \eqref{eq:path_integral} may give us the desired definition of quantum gravity. 

\begin{figure}[]
\begin{center}
\includegraphics[width=1\textwidth]{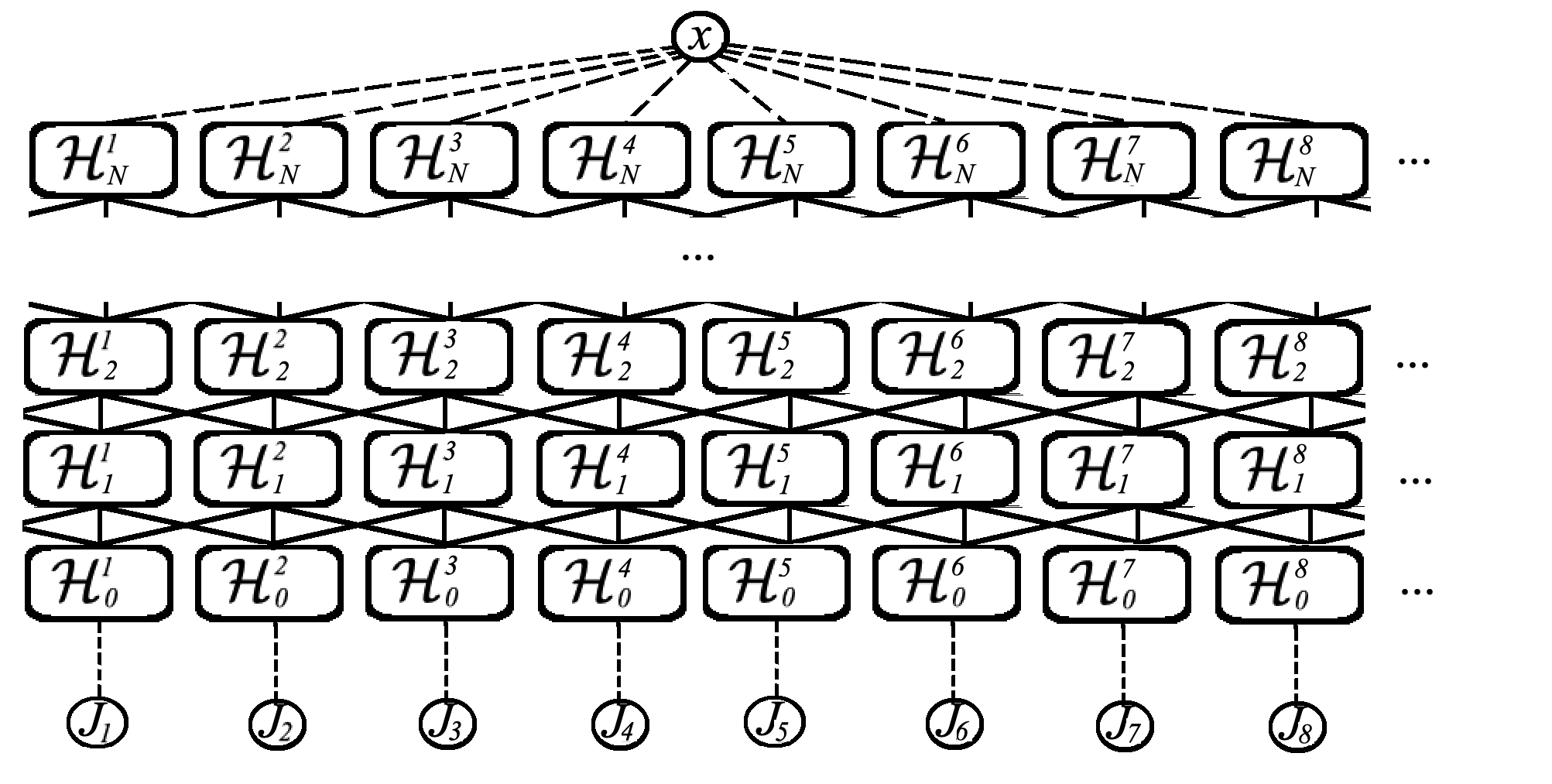}
\caption{Interactions network of a dual quantum field theory.} \label{fig:qft}
\end{center}
\end{figure}

To demonstrate how a local quantum field theory can emerge from a dual path integral, consider a collection of $n$'th order ``local'' operators $\hat{H}^{jk}_n$ which are non-zero only for $ j \in \{ k -1, k , k+1\}$ modulo $K$. (See Fig. \ref{fig:qft} for an illustration of interactions in a dual local quantum field theory with boxes representing separate Hilbert space ${\cal H}_n^k$ of the original quantum theory and solid lines representing local tensor products of operator from the respective factors.) If the non-zero operators satisfy an anti-commutation relation
\be
\{ \hat{H}^{ k, k+1}_n, \hat{H}^{ k, k-1}_n \} = 2  \hat{H}^{kk}_n = 2 \hat{I}^k_n \label{eq:commutator}
\ee
and a tracelessness condition
\bea
tr\left ( \hat{H}^{jk}_n\right ) = \delta_{jk} tr\left ( \hat{H}^{kk}_n  \right )  = \delta_{jk} tr\left (\hat{I}^k_n \right ) = \delta_{jk}{\cal N}^k_n \label{eq:traceless2}.
\eea
then the $n$'th order partition function is a straightforward generalization of Eq. \eqref{eq:anticomm_partition2},
\be
{\cal Z}^k_n(i x^k,  p_1, ..., p_K )  = {\cal N}^k_n e^{i x^k p_k} \cos\left( x^k \sqrt{p_{k-1}^2 +p_{k+1}^2} \right ).
\ee
The corresponding dual Hamiltonian density \eqref{eq:Hamiltonian_density} is complex
\be
{H}_n^k(x^k,  p_1, ..., p_K ) = i \log \left ( {\cal N}^k_n\cos \left (x^k \sqrt{ p_{k-1}^2 +p_{k+1}^2} \right ) \right )
\ee
 and as such cannot be interpreted as the standard Hamiltonian density of a local quantum field theory. However, recall that the variable $ix =i x^k_N$ of the fully interacting quantum partition function must be analytically continued to the real line $-\beta$ (see Sec. \ref{sec:continuation}). Then if we are only interested in the zero temperature limit, i.e. $\beta \rightarrow \infty$, the dominant contribution would come from only negative frequency mode, i.e.
\be
\cos \left (x_N^k \sqrt{(p^N_{k-1})^2 +(p^N_{k+1})^2}\right ) = \frac{e^{i x_N^k \sqrt{(p^N_{k-1})^2 +(p^N_{k+1})^2}} +  e^{- i x_N^k \sqrt{(p^N_{k-1})^2 +(p^N_{k+1})^2}}}{2} \rightarrow \frac{e^{- i x_N^k \sqrt{(p^N_{k-1})^2 +(p^N_{k+1})^2}}}{2}. \label{eq:cosine}
\ee
If we apply the same approximation, $- i x^k_n \rightarrow \infty$, to other factors of the partition function, then the local dual Hamiltonian density can be approximated as
\bea
{H}_n^k(x^k,  p_1, ..., p_K ) &\approx&  i \log \left ( {\cal N}^k_n/2 \exp \left ( - i x^k \sqrt{p_{k-1}^2 +p_{k+1}^2} \right ) \right ) \notag \\
 &=&   x^k \sqrt{p_{k-1}^2 +p_{k+1}^2} +  const
\eea
where the irrelevant constant does not produce any observable effects and can be dropped. As a result the dual Hamiltonian is the desired sum of the local dual Hamiltonian density terms 
\be
H_n[x^1, ..., x^K, p_1, ..., p_K] = \sum_{k=1}^K {H}_n^k(x^k,  p_1, ..., p_K ) = \sum_{k=1}^K x^k \sqrt{p_{k-1}^2 +p_{k+1}^2}\label{eq:dual_Hamiltonian}
\ee
as is the case for local quantum field theories. This puts the corresponding dual path integral in the same from as the Feynman path integral for a local quantum field theory on a lattice,
\be
{\cal  Z}[i x, J] \equiv  \int  \prod_{k=1}^K \frac{d x^k_0 d p_k^1  ... d p_k^N d x^k_N }{(2 \pi)^{N}} \;\Psi_\text{out}[x_N] e^{- \frac{i}{\hbar}  \sum_{n=1}^{N} \sum_{k=1}^K \left (x_n^k \sqrt{(p^n_{k-1})^2 +(p^n_{k+1})^2}  -  \left ( x^k_{n} -x^k_{n-1} \right )p_k^{n} \right )}  \Psi_\text{in}[x_0]  \label{eq:example3}
\ee 
where we inserted a ``dual'' Planck constant constant $\hbar=1$. Note that the limit $- i x^k_n \rightarrow \infty$ is equivalent to $i \hbar  \rightarrow 0$ and so only in this limit the dual path integral \eqref{eq:example3} represents the quantum system described in this section. On the other hand,  if we would have kept both modes in \eqref{eq:cosine}, which would be appropriate for example at finite temperatures, then the dual Hamiltonian would not be real and the standard quantum mechanical interpretation would be lost. This is the limit where we expect to see some non-trivial deviations from the Feynman path integral and where the effects of quantum gravity should become important. 

\section{Local interactions} \label{sec:spin}

In the previous sections we stated a possibility of using the dual path integral \eqref{eq:path_integral} to model or, more precisely, to define quantum gravity. This resonates well with the AdS/CFT proposal, but for the time being it also remains highly speculative. On a more practical level, it would be important to explore in greater details the strongly coupled quantum systems which can be solved using the dual description. In particular, are there any spin chain models whose partition functions can be expressed using the dual path integral representation developed in this paper? In this section we will discuss a couple of models for which the correlations function are local and in the following section we will discuss the Ising model of which correlation function are non-local. 

\begin{figure}[]
\begin{center}
\includegraphics[width=1\textwidth]{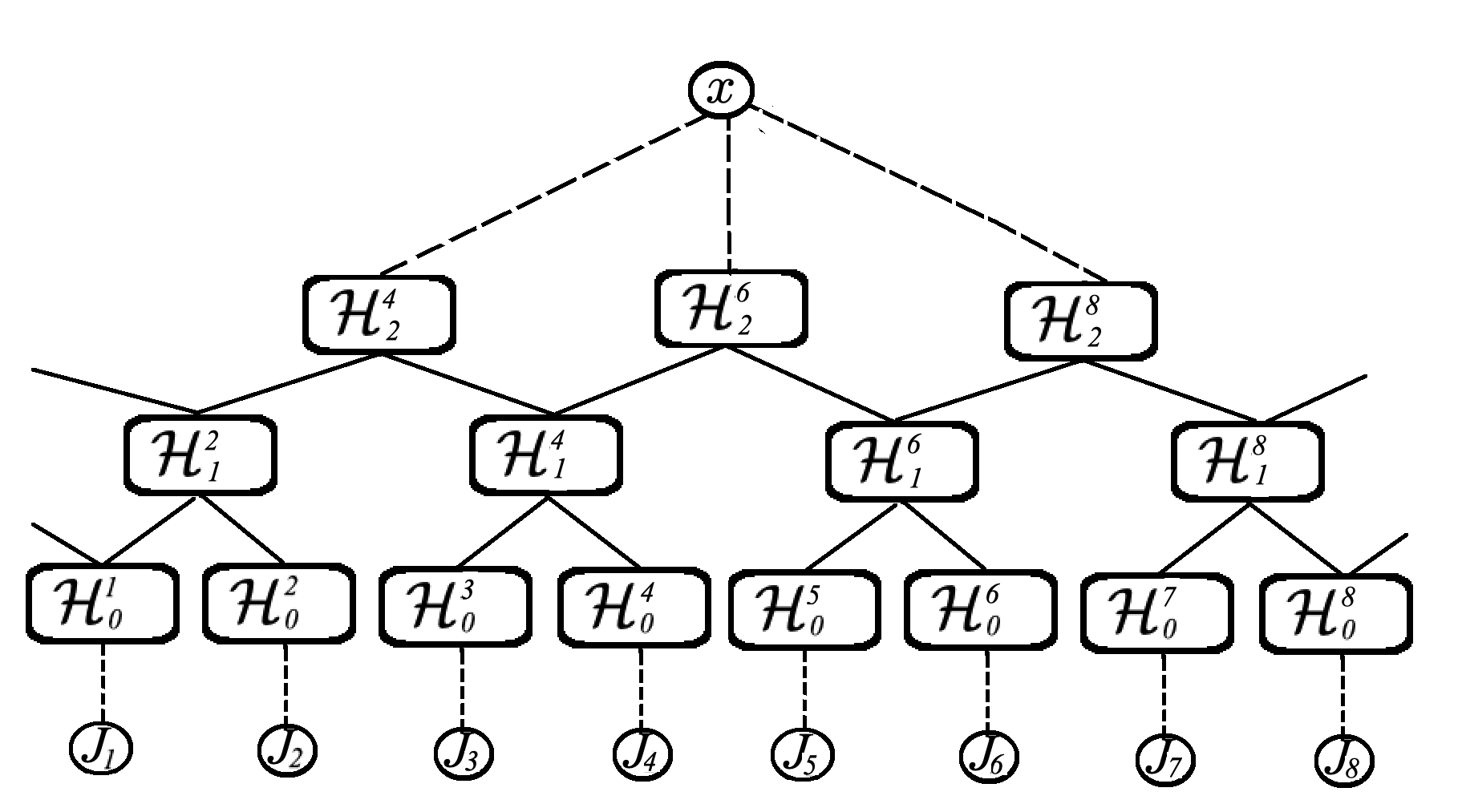}
\caption{Interactions network of a spin chain model.} \label{fig:tree}
\end{center}
\end{figure}

To construct a semi-simple example of such a model it is sufficient to consider a dual path integral with only two orders of interactions, i.e.  $N=2$. See Fig. \ref{fig:tree} for an illustration of interactions in the corresponding path integral with boxes representing factors of the Hilbert space ${\cal H}_n^k$ and solid lines representing tensor products of non-vanishing operator from the respective factors. As is evident from the figure, the only non-vanishing operators are $\hat{H}_1^{2k, 2k}, \hat{H}_1^{2k, 2k-1}, \hat{H}_2^{2k, 2k}, \hat{H}_2^{2k, 2k-2}$ which we assume to satisfy the following anti-commutation relations
\bea
\{\hat{H}_1^{2k, 2k}, \hat{H}_1^{2k, 2k-1}\} &=& 0\notag\\
\{\hat{H}_2^{2k,2k}, \hat{H}_2^{2k, 2k-2}\} &=& 0 \label{eq:relations}
\eea
where $K$ is even and all indices are periodic modulo $K$.  These operators can be expressed in terms of spin operators 
\bea
 \hat{H}_1^{2k, 2k-1} &=& \hat{\sigma}^x_{2k-1} \\
 \hat{H}_1^{2k, 2k} &=& \hat{\sigma}^y_{2k-1} \\
 \hat{H}_2^{2k, 2k-2} &=& \hat{\sigma}^x_{2k} \\
 \hat{H}_2^{2k, 2k} &=& \hat{\sigma}^y_{2k}
 \eea
 where $\hat{\sigma}^x_k$ and $\hat{\sigma}^y_k$ are the anti-commuting Pauli spin operators on lattice sites $k=1,...,K$. If we (for simplicity) also assume that 
 \be
  \hat{H}_0^k = \hat{I}_0
 \ee
then the corresponding quantum system can be written as a $1D$ spin chain model with Hamiltonian
\be
\hat{H} = \sum_{k=1}^{K/2} \left (J_{2k-1} \hat{\sigma}^x_{2k-1} + J_{2k} \hat{\sigma}^y_{2k-1}  \right ) \left ( \hat{\sigma}^x_{2k-2}  + \hat{\sigma}^y_{2k} \right ).
\label{eq:spin_model}
\ee

As a spin chain model the system of Hamiltonian \eqref{eq:spin_model} is strongly coupled (for generic $J_k$'s), but as a dual path integral it can be easily solved. All that we have to do is to plug in the known partition functions of the zeroth-, first- and second-order subsystems 
\bea
{\cal Z}_0^k[ix,p] &=& \exp(ixp) \label{eq:system0}\\
{\cal Z}_1^{2k}[ix,p_{1}, p_{2}] &=& \cos( x\sqrt{(p_{1})^2 + (p_{2})^2}) \\
{\cal Z}_2^{2k}[ix,p_{1}, p_{2}] &=&\cos( x\sqrt{(p_{1})^2 + (p_{2})^2})
\eea
into \eqref{eq:second_order}, i.e.
\bea
{\cal  Z}[ix, J]   &=& \int  \frac{\prod_{k=1}^K d x_0^k d p_{k}^1d \prod_{k=1}^{K/2}  x_1^{2k} d p_{2k}^2}{(2 \pi)^{3/2}}   e^{i\sum_{k=1}^K x_0^k (J_{k}- p_k^1)}\; \prod_{k=1}^{K/2} {\cal Z}^{2k}_1[ix^{2k}_1, p^1] e^{- i\sum_{k=1}^{K/2} x_1^{2k} p_{2k}^2} \; \prod_{k=1}^{K/2} {\cal Z}^{2k}_2[ix, p^2]\notag\\
 &=& \int  \prod_{k=1}^{K/2}\frac{d x^{k} d p_{k}}{2 \pi} \; \prod_{k=1}^{K/2} \cos\left ( x^{k} \sqrt{J_{2k}^2 + J_{2k-1}^2} \right ) e^{- i\sum_{k=1}^{K/2}  x^{k} p_{k}} \; \prod_{k=1}^{K/2} \cos\left ( x \sqrt{p_{k}^2 + p_{k-1}^2} \right ) .  
\eea
It is now straightforward to analytically continue $ix$ to $-\beta$ to obtain an exact expression for the fully interacting quantum partition function of a spin model \eqref{eq:spin_model},
\be
{\cal  Z}[-\beta, J] =  \int  \prod_{k=1}^{K/2}\frac{d x^{k} d p_{k}}{2 \pi} \; \prod_{k=1}^{K/2} \cos\left ( x^{k} \sqrt{J_{2k}^2 + J_{2k-1}^2} \right ) e^{- i\sum_{k=1}^{K/2}  x^{k} p_{k}} \; \prod_{k=1}^{K/2} \cosh\left ( \beta \sqrt{p_{k}^2 + p_{k-1}^2} \right ).\label{eq:example2}
\ee
Note that the integral of \eqref{eq:example2} has the exact form of the partition function \eqref{eq:first_order} with up-to first order interactions. This means that although we have started with the second order interactions, i.e. $N=2$, it was possible to reduce the order of interactions to  $N =1$. Upon integration with resect to $x^k$'s and $p_k$'s we obtain the following result
\bea
{\cal  Z}[\beta, J] &=& \prod_{k} \cosh\left ( \beta \sqrt{J_{2k}^2 + J_{2k-1}^2 + J_{2k-2}^2 + J_{2k-3}^2} \right )\label{eq:second_model} \\
&=& \prod_{k}\left ( 1 + \frac{\beta^2}{2}  \left ( J_{2k}^2 + J_{2k-1}^2 + J_{2k-2}^2 + J_{2k-3}^2 \right ) + \frac{\beta^4}{4!}\left ( J_{2k}^2 + J_{2k-1}^2 + J_{2k-2}^2 + J_{2k-3}^2 \right ) ^2 + ... \right ) .\notag
\eea
More generally we can consider the first-order interactions with  ${\cal Z}_0^k [ix_0^k, J_k] = \exp(ix_0^k J_k)$  and  ${\cal Z}_1^k [ix, p^1_{k+1}, ..., p^1_{k+L}] = \cos\left (x \sqrt{ \sum_{l=1}^L \left ( p^1_{k+l}\right )^2 } \right )$. Then the corresponding partition function is given by \eqref{eq:first_order} which can be written as
\bea
{\cal  Z}[\beta, J]  &=& \prod_{k} \cosh\left (\beta \sqrt{ \sum_{l=1}^L  J_{k+l}^2 } \right )\label{eq:third_model}
\\
&=& \prod_{k} \left (1+ \frac{\beta^2}{2} { \sum_{l=1}^L  J_{k+l}^2 } + \frac{\beta^4}{4!} \left ( { \sum_{l=1}^L  J_{k+l}^2 } \right )^2+ ...  \right ).\notag
\eea 

Correlation functions can be obtained by differentiating the partition functions with respect to the sources at the origin, i.e. at $J_k=0$ for all $k$. For the considered models the two-point correlation functions vanish and the smallest non-vanishing four-point correlation function is 
\be
\left \langle x_i^2 x_j^2 \right \rangle  \propto \left [  \frac{\partial^2}{\partial J_i^2}\frac{\partial^2}{\partial J_j^2} {\cal  Z}[\beta , J] \right ]_{J_k = 0}.
\ee
Evidently, calculation of the four-point correlation function amounts to calculating the respective terms in  \eqref{eq:second_model}  and \eqref{eq:third_model}. For example, if $i=j$, then 
\be
\left \langle x_i^4 \right \rangle \propto   \left [  \frac{\partial^4}{\partial J_i^4} {\cal  Z}[\beta , J] \right ]_{J_k = 0}  =   \left (\left ( \frac{\beta^2}{2} \right )^2+ 2 \frac{\beta^4}{4!} \right ) 4!  =  8\beta^4
\ee
for the model described by Eq.  \eqref{eq:second_model}  and 
\be
\left \langle x_i^4 \right \rangle \propto   \left [  \frac{\partial^4}{\partial J_i^4} {\cal  Z}[\beta , J] \right ]_{J_k = 0}  =   \left (\frac{ L (L-1)}{2} \left ( \frac{\beta^2}{2} \right )^2+ L  \frac{\beta^4}{4!} \right ) 4!  =   (3 L^2 - 2L ) \beta^4
\ee
for the model described by Eq.  \eqref{eq:third_model}. By conducting similar calculations for arbitrary $i$ and $j$ we get the following results
\bea
\left \langle x_i^2 x_j^2 \right \rangle \propto \begin{cases} 
\left (\left ( \frac{\beta^2}{2} \right )^2+ 2 \frac{\beta^4}{4!} \right ) 4!  =  8 \beta^4   \;\;\; \;\;\; \;\;\;\; \;\;&\text{for}\;\;i=j\\
\left (2 \left ( \frac{\beta^2}{2} \right )^2+ 4 \frac{\beta^4}{4!} \right ) 2! 2!  = \left ( 2+ \frac{2}{3} \right ) \beta^4    \;\;&\text{for}\;\;i=2k \;\;\text{and}\;\; j = 2k - 1\\
\left (3 \left ( \frac{\beta^2}{2} \right )^2+ 2 \frac{\beta^4}{4!} \right ) 2! 2!  = \left ( 3 + \frac{1}{3} \right )   \beta^4    \;\;&\text{for}\;\;i=\{ 2k, 2k-1\} \;\;\text{and}\;\; j \in \{ 2k -2, 2k-3\}\\
\left (4 \left ( \frac{\beta^2}{2} \right )^2 \right ) 2! 2!  = 4 \beta^4     \;\;\; \;\; \;\; \;\;\;\; \;\; &\text{otherwise}
\end{cases}\notag
\eea
and
\bea
\left \langle x_i^2 x_j^2 \right \rangle \propto \begin{cases} 
 \left (\frac{ L (L-1)}{2} \left ( \frac{\beta^2}{2} \right )^2+ L  \frac{\beta^4}{4!} \right ) 4!  =  (3 L^2 - 2 L) \beta^4   \;\;\; \;\;\; \;\;\;\; \;\;\;\; \;\;\;\; \;\;\;\;&\text{for}\;\;i=j\\
 \left (\left ((2 L -  | i-j |) | i-j | \right )\left ( \frac{\beta^2}{2} \right )^2+  (2 (L - | i-j |) ) \frac{\beta^4}{4!} \right ) 2!2! = \\\;\;\;\;\;\;\;\;\; \;\;\;\;\;\;\;\;\;\;\;\;\;\;\;\;\;\;\;\;\;\;\;\;\;\;\;\;\;\;\;\;\;\;=  \left ( (2 L -  | i-j |) | i-j |  + \frac{L-| i-j |}{3}  \right ) \beta^4    \;\; \;\;\;\;&\text{for}\;\;1 \le | i-j | \le L \\
 \left ( L^2 \left ( \frac{\beta^2}{2} \right )^2 \right ) 2! 2!  =  L^2 \beta^4   \; \;\;\;\; \;\;\;\; \;\;\;\; \;\;\;\; \;\;\;\; \;\; \;\;  \;\;\;\; \;\;\;\;\;\;&\text{for} \;\; | i-j | > L 
\end{cases} \notag
\eea
for \eqref{eq:second_model} and \eqref{eq:third_model} respectively.  It is interesting to note that for the latter model the correlation function takes the largest value at $| i-j |=0$, then quickly drops to the smallest value at $| i-j |=1$ and then gradually grows to an asymptotic value at $| i-j | \ge L$. Of course, given a fully non-perturbative expression for the partition functions it should not be too surprising that we are able to quickly obtain the obsevables. Although we only calculated the four-point correlation functions, but it is a straightforward exercise to calculate an arbitrary high-order correlation function by differentiating the respective partition functions with respective sources.

\section{Ising model}\label{sec:ising}

In the previous section we discussed some semi-simple spin chain models described by partition functions \eqref{eq:second_model}  and \eqref{eq:third_model} with up to first order interactions, but it turned out that the correlations were completely local. On the other hand in the most general expression of the dual path integral \eqref{eq:path_integral} the higher order interactions are intrinsically non-local.  Then one might wonder if non-locality is something that is only present in theories with higher order interactions, or if it is possible to construct a non-local theory with, for example, only local first order interactions or, in other words, using only local first order systems. 

\begin{figure}[]
\begin{center}
\includegraphics[width=1\textwidth]{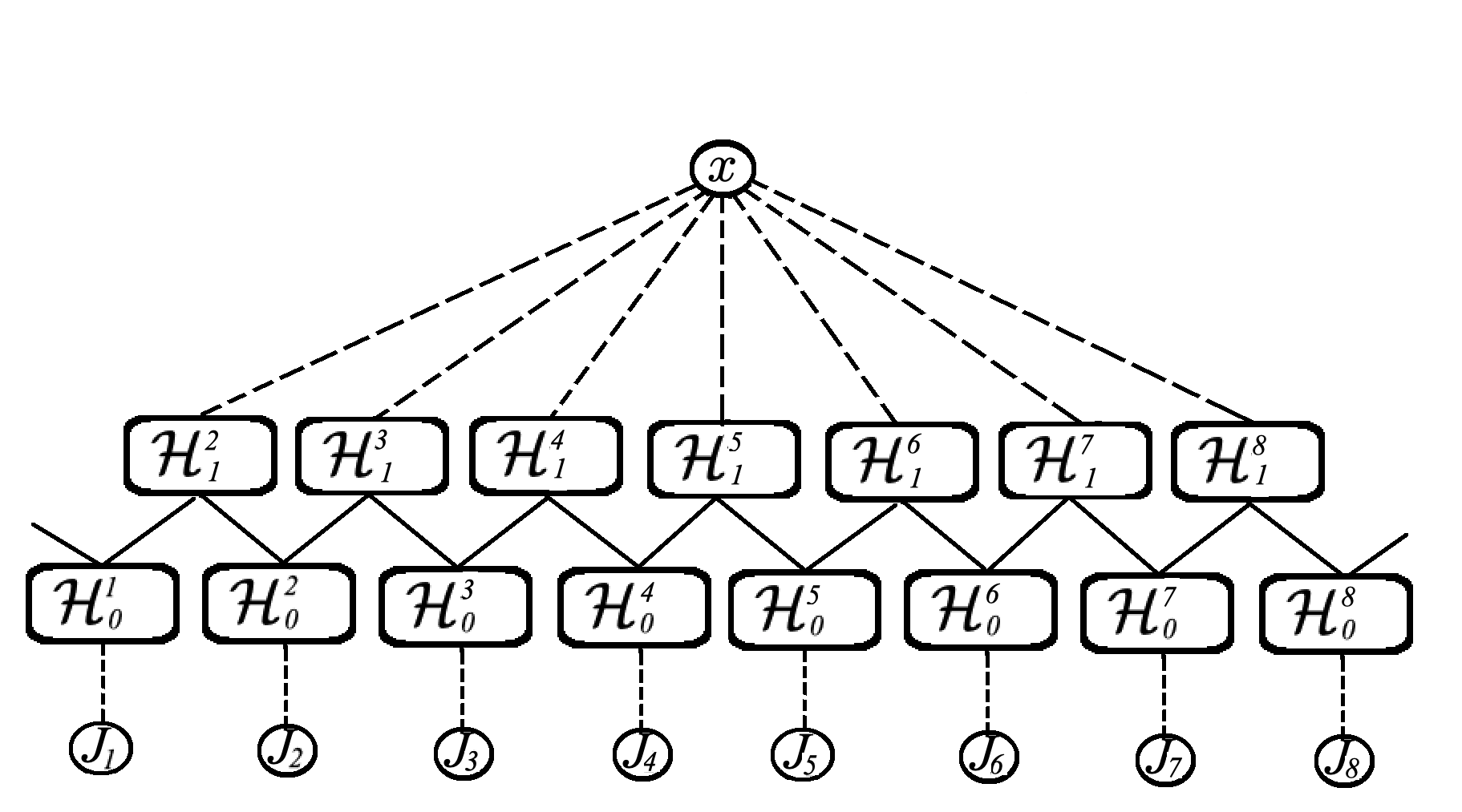}
\caption{Interactions network for Ising model.} \label{fig:ising}
\end{center}
\end{figure}

Consider a partition function  \eqref{eq:first_order}  described by the zeroth order systems with partition functions,
\be
{\cal Z}_0^k[i x_0^k, J_{k}]  =  \cos(x_0^k J_{k}) = \frac{1}{2} \left ( \exp(ix_0^k J_{k}) + \exp(- ix_0^k J_{k})\right )
\ee
and by the first order systems with local and homogeneous partition functions, 
 \be
{\cal Z}^j_1[ix, J_{1}, ..., J_{K}] =  {\cal Z}_1[ix, J_{j}, ..., J_{j+L}].
 \ee
 Here the locality is represented by  the assumption $1 \le L\ll K$ and the homogeneity by the assumption ${\cal Z}_1^j= {\cal Z}_1$.  Then the partition function \eqref{eq:first_order} can be expressed as a product of the first order partition functions
\bea
{\cal  Z}[ix, J]   &=&  \int  \prod_{k=1}^K\frac{d x_0^k d p_k^1}{2 \pi}    \prod_{k=1}^K {\cal Z}_0^k[i x_0^k, J_{k}]\; e^{- i\sum_{k=1}^K x_0^k p_k^1}\; \prod_{j=1}^K {\cal Z}^j_1[ix, p^1_1, ..., p^1_K]\notag\\
&=&  \int  \prod_{k=1}^K\frac{d x_0^k d p_k^1}{2 \pi}    \prod_{k=1}^K \; \left(  e^{ - i\sum_{k=1}^K x_0^k ( p_k^1 - J_{k})}+e^{ - i\sum_{k=1}^K x_0^k ( p_k^1+J_{k} )}\right ) \prod_{j=1}^K {\cal Z}^j_1[ix, p^1_1, ..., p^1_K]\notag\\
&=&  \sum_{s_k \in \{-1,+1\}}  \;\;\;\; \prod_{j=1}^K {\cal Z}^j_1[ix, s_1 J_{1}, ..., s_K J_{K}] \notag\\
 & =&  \sum_{s_k \in \{-1,+1\}}  \;\;\;\; \prod_{j=1}^K {\cal Z}_1[ix, s_{j} J_{j}, ..., s_{j+L} J_{j+L}].\label{eq:nonlocal1}
\eea
For example, if we set the first order partition function to be
\be
{\cal Z}_1[\beta, A,B] = \exp \left ( \beta A B + \beta \frac{A + B}{2}  \right ),
\ee
i.e. $L=1$, then
\be
{\cal  Z}[\beta, J] =  \sum_{s_k \in \{-1,+1\}}  \;\;\;\; \prod_{j=1}^K \exp \left ( \beta  s_j J_{j} s_{j+1} J_{j+1}  + \frac{\beta}{2} \left ( s_{j} J_{j}  +  s_{j+L} J_{j+L} \right )  \right ) .\label{eq:nonlocal2}
\ee
(See Fig. \ref{fig:ising}  for an illustration of interactions in the corresponding path integral.)
Furthermore if we set all sources to be equal, i.e. $J_j = J$, then the partition function \eqref{eq:nonlocal2} can be rewritten as 
\be
{\cal  Z}[\beta, J] =  \sum_{s_k \in \{-1,+1\}}  \;\;\;\; \prod_{j=1}^K T(s_j, s_{j+1}).\label{eq:Ising_part}
\ee
or in a more standard form
\be
{\cal  Z}[\beta, J] = Tr \left ( T^K \right ) = \lambda_+(\beta, J)^K + \lambda_-(\beta, J)^K  \label{eq:Ising_trace}
\ee
where the trace is with respect to two-by-two matrix obtained after raising to $K$'th power the so-called ``transfer matrix'' 
\be
T = \begin{pmatrix}
T(+1, +1) \;\;& \;\;T(+1, -1)   \\
T(-1, +1) \;\; & \;\; T(-1, -1)  
\end{pmatrix}  = \begin{pmatrix}
\exp \left (\beta J^2 + \beta J\right ) \;\;\;\; &  \exp \left (- \beta J^2 \right ) \;\;\;\;\;\;  \\
\exp \left (- \beta J^2 \right ) \;\;\;\; \;\;\;\; & \exp \left (\beta J^2 - \beta J\right )
\end{pmatrix}.\label{eq:transfer_mat}
\ee
and $\lambda_\pm(\beta, J)$ are the two eigenvalues of the matrix.  

Equation \eqref{eq:transfer_mat} has the exact form of the transfer matrix for 1D Ising model described by Hamiltonian
\bea
\hat{H} &=& \sum_{k=1}^{K} \left (J^2 \hat{\sigma}^x_{k}\hat{\sigma}^x_{k+1} + J \hat{\sigma}^x_{k}  \right ) \notag\\
&=& \sum_{k=1}^{K} \left (J^2 \hat{\sigma}^x_{k}\hat{\sigma}^x_{k+1} + J \frac{\hat{\sigma}^x_{k} + \hat{\sigma}^x_{k+1} }{2} \right ).
\label{eq:Ising_model}
\eea
The corresponding partition function ${\cal  Z}[\beta, J]$ has only two free parameters $\beta J$ and $\beta J^2$, which in our case is parametrized by $\beta$ and $J$. This shows that the dual description can be used to study the strongly coupled systems including certain spin chain models such as Ising model. 

\section{Discussion}\label{sec:discussion}

In this paper we simultaneously achieved two parallel results: derived a dual path integral representation of some strongly coupled  systems  \eqref{eq:path_integral} and then argued that the dual description may be responsible for the emergence of space-time, quantum field theories and gravity. While the first result is purely mathematical and should be viewed as a non-perturbative method for calculating partition functions, the second result is an attempt to study how the essential ingredients of any successful theory of gravity may emerge from a dual path integral representation of the partition function for strongly coupled systems. With this respect our approach is similar to the AdS/CFT  with the main difference that we do not a priory assume a specific symmetry of the interactions (e.g. conformal) nor a specific geometry of the dual space-time (e.g. anti-de Sitter) and rely completely on the interactions to determine the geometry of the dual space-time. Our approach is also similar to the more recent ideas of describing space-time using entanglement between subsystems \cite{Ryu, VanRaamsdonk, Swingle, Carroll, Vanchurin4} or using quantum circuits \cite{Almheiri, Susskind, Vanchurin5}, but our derivation of the dual description is certainly very different. 

To demonstrate how the non-perturbative method works in practice, we considered three examples of quantum mechanical systems with up to the first-order interactions complexity (see Sec. \ref{sec:spacetime}, \ref{sec:spin} and \ref{sec:ising}) and one example with arbitrary high-order  interactions complexity (see Sec. \ref{sec:qft}) and calculated the respective dual expressions for the fully interacting partition functions. In the latter example we showed that the zero temperature limit of the dual path integral can be approximated as a path integral of a $1+1$D quantum field theory with a non-canonical kinetic term and with the orders of interactions playing the role of a dual time. In this limit the dual system is a legitimate quantum field theory on a lattice, but for more general systems we expect the dual path integral to differ from the Feynman path integral. This is where our analysis diverges significantly from the AdS/CFT proposal in which one is supposedly dealing with legitimate quantum theories on both sides of the duality. Of course, in a true theory of quantum gravity the Feynman path integral may not be adequate and with this respect the dual path integral could give us the desired non-perturbative definition of quantum gravity.

To study the emergent phenomena we first followed the analysis of Ref. \cite{Vanchurin2} and defined an extended partition function (Sec. \ref{sec:spacetime}) which solves a relativistic wave equation and simultaneously describes both non-interacting and interacting partition functions up to the first order in interactions complexity. Although the original system was in a time-invariant thermal state (Sec. \ref{sec:interactions}), in the dual description the relativistic dynamics emerges from interactions between subsystems. With this respect not only time, but also space-time can emerge as was first noted in Ref. \cite{Vanchurin2}. Unfortunately, such systems are not rich enough to describe many strongly coupled systems and therefore not suitable for describing the more complex emergent phenomena such as quantum field theories, curved space-time or gravity. To study the more general emergent phenomena we proposed to use the dual path integral representation of the partition functions with arbitrary high orders of interactions. The higher-order interactions are expected in a generic strongly coupled system and in Sec. \ref{sec:qft} we defined one such system using the interactions network of Fig. \ref{fig:qft}. As was already mentioned, the corresponding dual path integral can be approximated as the Feynman path integral of a $1+1$D quantum field theory on a flat background,  but a generalization of Fig. \ref{fig:qft} to more general curved backgrounds does not seem out of reach.  Also note that it is not too difficult to come up with examples of different, but equivalent, dual path integral representations of the same quantum system and so at some level the emergence of local symmetries is expected. What is, however,  less obvious is how to derive fully non-perturbative equations of general relativity, but that is work in progress. 

{\it Acknowledgments.} The work was supported in part by Foundational Questions Institute (FQXi).

{\it Data Availability Statement. }The data that support the findings of this study are available from the corresponding author upon reasonable request.

\end{document}